\documentclass[final,3p,times]{elsarticle}

\usepackage{graphicx}
\usepackage{multirow}
\usepackage{natbib}
\usepackage{hyperref}
\usepackage{subfig}
\usepackage{xcolor}
\usepackage{amssymb}

\usepackage{amssymb}

\journal{Nuclear Physics A}

\begin{document}

\begin{frontmatter}


\title{Investigation of reaction and $\alpha$ production cross sections with $^9$Be projectile}
\author[a,b]{Satbir~Kaur}
\author[a,b]{V.~V.~Parkar} \ead{Corresponding author: vparkar@barc.gov.in}
\author[a,b]{S.~K.~Pandit}
\author[a,b]{A.~Shrivastava}
\author[a,b]{K.~Mahata}
\author[a]{K.~Ramachandran}
\author[a,b]{Sangeeta~Dhuri}
\author[a,b]{P.~C.~Rout}
\author[a]{A.~Kumar}
\author[a,b]{Shilpi~Gupta} 
\address[a]{Nuclear Physics Division, Bhabha Atomic Research Centre, Mumbai - 400085, India}
\address[b]{Homi Bhabha National Institute, Anushaktinagar, Mumbai - 400094, India}

\begin{abstract}
In order to investigate the contribution of $\alpha$ production in the reaction cross sections, measurements of elastic scattering and inclusive $\alpha$ particle angular distributions have been carried out with the $^9$Be projectile on $^{89}$Y, $^{124}$Sn, $^{159}$Tb, $^{198}$Pt, and $^{209}$Bi targets over a wide angular range at energies near the Coulomb barrier.  The measured elastic scattering angular distributions were fitted with optical model calculations, and reaction cross sections were extracted.  The same data were also analysed using both global optical model potentials (Global OMP) and microscopic S$\tilde{a}$o Paulo potentials (SPP), to obtain the reaction cross sections. The data available in the literature for $^9$Be projectile includes the elastic scattering angular distributions, $\alpha$ production cross sections, and complete fusion cross sections on these and other targets at several energies are also utilised for comparative studies. The reaction cross section extracted from the three potentials (Best Fit, Global OMP and SPP) are in reasonable agreement for all the targets except for the energies below the barrier where the results from SPP deviate by 30-50 \%. Inclusive $\alpha$ particle production cross sections were also extracted by integrating the $\alpha$ particle angular distributions. The present data and data available from literature of reaction and $\alpha$-particle production cross sections were utilised to make systematic studies. Systematics of reaction and $\alpha$-particle production cross sections revealed their universal behaviour. The ratio of complete fusion (CF), inclusive $\alpha$ cross sections, and their sum (CF+$\alpha$) to reaction cross sections show that at sub-barrier energies inclusive $\alpha$ dominates and above barrier CF dominates, and cumulative of these two processes almost explains the reaction cross section at all energies over a wide range of target mass. A comparative study of $\alpha$ production with other weakly bound projectiles is also performed and a clear distinction between projectile types is observed, which is found to correlate well with $\alpha$ separation energies of the projectiles.
\end{abstract}


\end{frontmatter}


\section{\label{sec:Intro}Introduction}

In recent times, study of nuclear reactions with nuclei having a predominant weakly bound $\alpha$+x cluster structure ($^{6,7}$Li, $^9$Be), have renewed interest to probe the origin of large $\alpha$ production cross sections and the role of coupling to low-lying states in the continuum on various reaction channels \cite{Jha20, Kolata16, CANTO15, KEELEY09}. For understanding the mechanism of large $\alpha$ production, detailed measurements of energy and angular distributions as well as integrated inclusive $\alpha$ cross sections have been investigated with $^{6,7}$Li projectiles \cite{Signorini03, Pakou03, Pakou05, Pako06, Shri06, Pako07, Souz09, Sant09, Kumawat10, Souza2010, Santra12, Luong13, Pand16, Pandit17, Chatto16, Chatto18, chavi}. Complete and incomplete fusion as well as exclusive particle coincidence measurements along with detailed calculations have been performed to investigate the origin of large $\alpha$ production cross sections \cite{Signorini03, Pakou03, Pakou05, Pako06, Shri06, Pako07, Souz09, Sant09, Kumawat10, Souza2010, Santra12, Luong13, Pand16, Pandit17, Chatto16, Chatto18, chavi, Lei15, Lei17, sanat21, kcook19}.  Recent studies with $^7$Li have revealed the dominance of direct $t$-stripping mechanism over breakup fusion for large $\alpha$ production~\cite{sanat21, kcook19}. The systematics and mechanisms of $\alpha$ particle production with weakly and strongly bound projectiles have also been reported recently \cite{VVP_EPJA23}. 

 In the case of $^9$Be, there are several elastic scattering measurements available in the literature~\cite{Umbelino22, Harphool22, Arazi18, Morales16, Palshetkar, Zamora11, Di10, Yu_2010, GomesSm209, Benjamim07, GomesSm06, GVM05, PRS05, Gomes04,Signorini2000, JARCZYK81, Bodek80, JsEck80, Balzer77, DP77}. From the elastic scattering measurements, optical model potentials and reaction cross sections were extracted. The global optical model potential for the $^9$Be projectile was also developed using the available elastic scattering angular distribution data \cite{Yongli}. Although there are several elastic scattering measurements, very few measurements of $\alpha$ particle angular distributions and hence its production cross section have been reported  ~\cite{Harphool22, Palshetkar, Bodek80, woolli01, Sig04}. It will be interesting to study the contribution of $\alpha$ production in the reaction cross section to understand the complete reaction mechanism. In the studies with $^{6,7}$Li projectiles, it is demonstrated that the $\alpha$ production along with complete fusion (CF) cross sections almost exhausts the reaction cross sections \cite{Jha20, Santra12, Pandit17}. However, similar study does not exist with $^9$Be projectile. Due to the low neutron as well as $\alpha$ separation energies and also the Borromean cluster structure in $^9$Be \cite{Pandit11}, the mechanism of $\alpha$ production can be different as compared to $^{6,7}$Li projectiles. Neutron transfer assisted breakup process could be a significant part of $\alpha$ production for projectiles having low neutron separation energies like $^9$Be \cite{Rafi10, Cook2016, Papk07, Brow07}.  

The present work aims to study the contribution of inclusive $\alpha$ production cross section to the reaction cross section for $^9$Be induced reaction, for a wide range of targets, at energies around the Coulomb barrier. The reaction cross section can be estimated from the optical model analysis of the elastic scattering angular distributions. Further, the potential parameters obtained using elastic scattering data could be an important input to entrance channel potential in the coupled channel calculations. Hence, these potential parameters can be used to investigate the influence of different inelastic, transfer, and breakup channels on elastic and fusion cross sections and to estimate the cross sections of these non-elastic channels as well. 

In this work, we have measured elastic scattering angular distributions with $^9$Be projectile on various targets ($^{89}$Y, $^{124}$Sn, $^{159}$Tb, $^{198}$Pt, and $^{209}$Bi) to extract the reaction cross sections and the angular distribution of the $\alpha$ particles to obtain the inclusive $\alpha$ production cross sections. A systematic study has been performed using present data along with available literature data for $\alpha$ production and complete fusion. The paper is organized as follows: The experimental details are described in Sec.\ \ref{sec:exp}. The measured elastic scattering angular distributions along with the optical model calculations using Global OMP, SPP are discussed in Sec.\ \ref{sec:OMA}. The measured energy spectra and inclusive $\alpha$ particle angular distributions are reported in Sec.\ \ref{sec:alphaenergy}. Systematic studies of the reaction cross sections, CF and $\alpha$ production cross sections with $^9$Be projectile are described in Sec.\ \ref{sec:systematics}. The summary is given in Sec.\ \ref{sec:summary}.

\section{Experimental Details} \label{sec:exp}
\begin{figure}
\centering
\includegraphics[width=100mm]{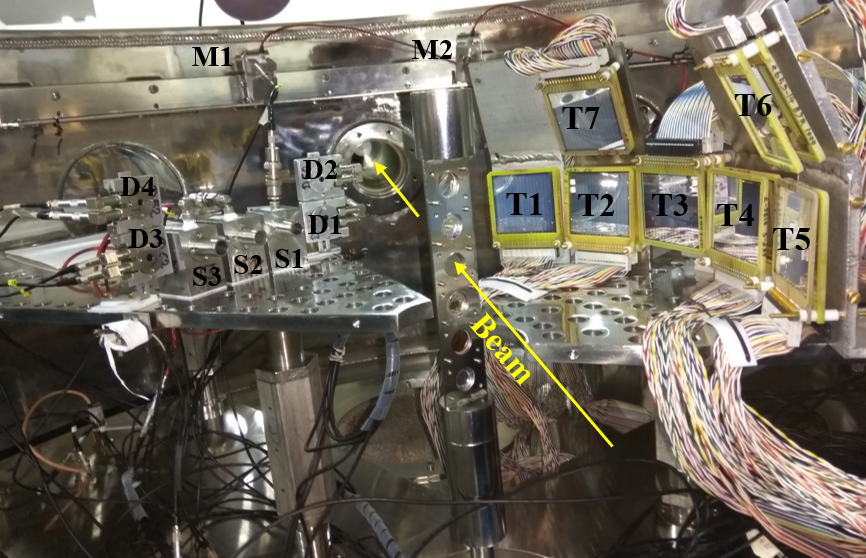}
\caption{\label{exptl_setup} Photograph of the experimental setup consisting of seven Si-strip detector telescopes (T1-T7), three silicon surface barrier (SSB) detector telescopes (S1-S3), and four Si-integrated detector telescopes (D1-D4) inside a 1.5 m diameter scattering chamber. Two SSB detectors (M1, M2) kept at $\pm$20$^{\circ}$ for absolute normalization and beam monitoring are also denoted.}
\end{figure}

\begin{figure}
\centering
 \includegraphics[width=100mm]{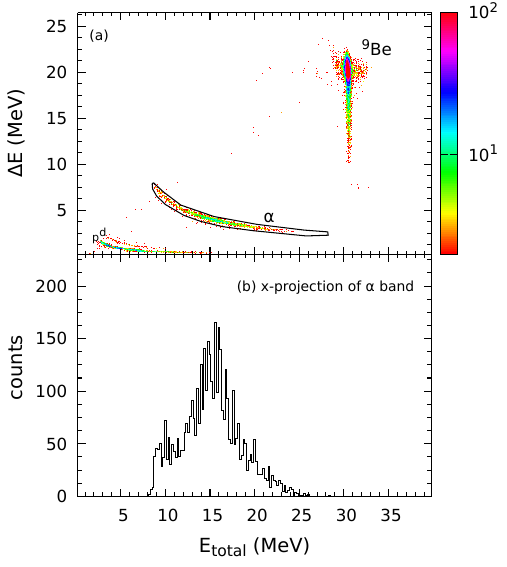}
\caption{\label{2Dspectra} (a) Typical two-dimensional energy calibrated spectrum of $\Delta$E vs E$_{total}$ for $^9$Be+$^{159}$Tb reaction at energy E$_{lab}$=35.3 MeV and $\theta_{lab}$ = 86$^{\circ}$. (b) x-projection of $\alpha$ band. }
\end{figure}

The measurements were performed using the $^9$Be beam at the BARC-TIFR Pelletron-LINAC Facility, Mumbai, India. The list of self-supporting targets used along with beam energies are given in Table\ \ref{tab:targetdetails}.  Beam energies were corrected for the energy loss at half the target thickness and were further used in the analysis. The corrected beam energies are denoted as E$_{lab}$ in the table. The corresponding E$_{c.m.}$/V$_B$ values are also listed where, V$_B$ values are estimated from Ref.\ \cite{NRV}. A photograph of the experimental setup is shown in Fig.\ \ref{exptl_setup}. As can be seen, the experimental setup consisted of seven Si-strip detector (Micron Semiconductors make) telescopes (T1-T7), three silicon surface barrier (SSB) detector (ORTEC make) telescopes (S1-S3), four Si-integrated detector (monolithic silicon $\Delta$E-E) telescopes (D1-D4) (developed by Electronics Division, BARC, Mumbai \cite{Topkar11}) along with two monitor detectors (M1,M2) inside a 1.5 m diameter scattering chamber. For the Si-strip detector  telescopes (T1-T7), the $\Delta$E detectors were single-sided with 16 vertical strips, and the E detectors were double-sided with 16 strips on front (vertical) and 16 strips on the back (horizontal). Typical thicknesses of $\Delta$E strip detectors were 50 $\mu$m (T1, T2, T5, T6 and T7) and 20 $\mu$m (T3, T4). While the E strip detectors were of thicknesses 1.5 mm (T1-T5 and T7) and 1 mm (T6). The active area of the Si-strip detectors was 50 x 50 mm$^2$ and the width of each strip was 3.1 mm. For SSB detectors (S1-S3), the thicknesses of the $\Delta$E detectors were 45 $\mu$m (S1, S2) and 25 $\mu$m (S3). In the case of E detectors, the thicknesses were 1mm (S2, S3) and 5 mm (S1). For the Si-integrated detectors (D1-D4), the thicknesses were 15 $\mu$m and 300 $\mu$m for $\Delta$E and E layers, respectively. Two SSB detectors (M1, M2) with thicknesses $\sim$ 300 $\mu$m were kept at $\pm$20$^{\circ}$ for absolute normalization and beam monitoring. The angular range covered with the detector setup was from $\theta_{lab}$ = 16$^\circ$ to 160$^\circ$. The data were collected in an event-by-event mode using a VME data acquisition system. The triggers for the strip telescopes (T1-T7) were taken from the Mesytec make MSCF16 modules of the E detectors, which correspond to "OR" of all 16 strips. Final master gate was generated as the "OR" of the triggers from the strip telescopes, E of SSB telescopes and monitor detectors. The detectors were calibrated using the energy of $\alpha$ particles emitted from the $^{229}$Th source and the elastic scattering data from different targets listed above. A typical two-dimensional energy calibrated spectrum of $\Delta$E vs E$_{total}$ for $^9$Be+$^{159}$Tb system at E$_{lab}$=35.3 MeV and $\theta_{lab}$ = 86$^{\circ}$ is shown in Fig.\ \ref{2Dspectra}(a). The x-projection of the $\alpha$ band is shown in Fig.\ \ref{2Dspectra}(b). 

\begin{table}

\caption{\label{tab:targetdetails} Details of targets used, beam energies, corrected lab energies at half thickness of the target, and E$_{c.m}$/V$_B$ values. The V$_B$ values used are calculated from Ref.\ \cite{NRV}.}
\centering
\begin{tabular}{cccccc}
\hline \hline
&&&&&\\
Target&thickness &E$_{beam}$&E$_{lab}$&E$_{c.m}$/V$_B$\\
& (mg/cm$^2$) &(MeV)&(MeV)&\\
&&&&&\\
\hline
&&&&&\\
$^{209}$Bi&0.5&35.8&35.7&0.85\\
$^{198}$Pt&1.5&35.8&35.4&0.89\\
$^{159}$Tb&1.7&35.8&35.3&1.02\\
$^{124}$Sn&1.7&29.7&29.1&1.03\\
&&35.8&35.2&1.25\\
$^{89}$Y&1.0&29.7&29.3&1.24\\
&&&&&\\
\hline \hline
\end{tabular} 
\end{table}

\section{Results and Discussion}
\subsection{\label{sec:OMA} Elastic scattering angular distributions}

\begin{figure}
\centering
\includegraphics [width=100mm] {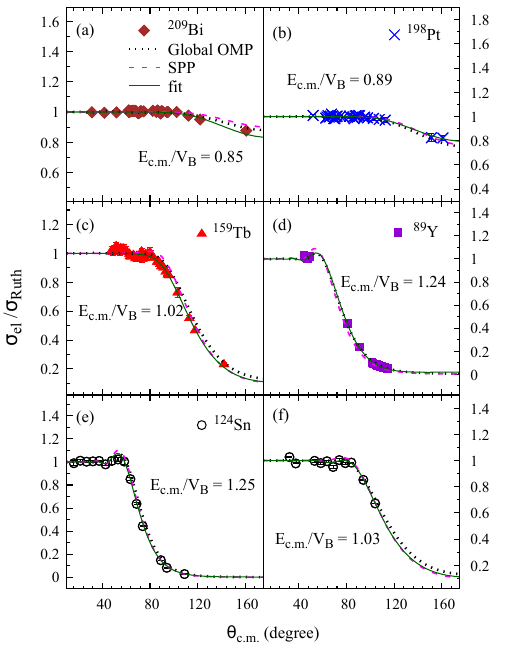}
\caption{\label{All_elas} Elastic scattering angular distributions for (a)$^{209}$Bi, (b) $^{198}$Pt, (c) $^{159}$Tb, (d) $^{89}$Y, (e,f) $^{124}$Sn targets. The solid lines are the results of the optical model fit, the dotted lines represents the optical model calculations using global optical model potentials and the dashed lines represent the calculations with SPP.}
\end{figure}




The experimental elastic scattering cross sections normalized to the Rutherford cross sections for the $^9$Be projectile on $^{89}$Y, $^{124}$Sn, $^{159}$Tb, $^{198}$Pt, and $^{209}$Bi targets are shown in Fig.~\ref{All_elas}. \textcolor{black}{The cross section of the corresponding reaction channels (elastic scattering or $\alpha$ particle) was obtained by taking the ratio of the yields of the corresponding reaction channel (elastic scattering or $\alpha$ particle) to the yield of the monitor detector. In this method,} the uncertainties arising due to the target thickness and beam current were eliminated. \textcolor{black} {The ratio of the two monitor counts was found to be around 1 throughout the run. Thus, the effect of beam wandering was not significant.} Hence, the error bars in the cross sections are due to statistical errors only. Optical model (OM) calculations were performed using both the phenomenological optical model potentials \cite{Yongli} and microscopic S$\tilde{a}$o Paulo potentials \cite{SPP2, SPP1} with the code {\footnotesize FRESCO} (version {\footnotesize FRES} 2.9) \cite{Fresco}. For phenomenological optical model potential, the recently developed global optical model potential (Global OMP) for the $^9$Be projectile \cite{Yongli} was used. This optical potential has a Woods-Saxon form: V(r,E) = V$_{R}$(r,E) + i[W$_{S}$(r,E)+W$_{V}$(r,E)] + V$_{C}$(r) where V$_{R}$ denotes the real part of the potential, W$_{S}$ and W$_{V}$ are the surface and volume imaginary parts of potential, respectively, and V$_{C}$(r) is the Coulomb potential. Based on this potential, the energy-dependent parameters defined in Ref.\ \cite{Yongli} were used for the calculations.
The calculations using the global optical model potential are shown as dotted lines in Fig.~\ref{All_elas}.  
The optical model calculations were also performed using the microscopic double-folding S$\tilde{a}$o Paulo potential (SPP), which takes into account the effects of Pauli non-locality related to the exchange of nucleons between the projectile and target \cite{SPP2,SPP1}. In the context of OM analysis (V$_{OM}$ = V$_{SPP}$ (N$_R$+iN$_I$)), the imaginary potential has the same form as the real part with some normalization constant. For real and imaginary parts of the optical potential, the strength coefficients are N$_R$ = 1.0 and N$_I$ = 0.78. This procedure has been shown to be suitable for describing the elastic scattering cross sections for many systems in wide energy range \cite{Gasq06}. The calculations using SPP are shown \textcolor{black} {with} dashed lines in Fig.~\ref{All_elas}.  The elastic scattering angular distributions were also fitted using the {\footnotesize SFRESCO} module of the {\footnotesize FRESCO} code \cite{Fresco}, utilizing Woods-Saxon forms for both real and imaginary potentials. The calculations using {\footnotesize SFRESCO} are shown as solid lines in Fig.~\ref{All_elas}.  The extracted reaction cross sections from all three (Best fit, Global OMP and SPP) different calculations are given as $\sigma_R$ (Ref.), $\sigma_R$ (Global OMP) and $\sigma_R$ (SPP) respectively in Table~\ref{tab:table2}.  The available elastic scattering data for $^{28}$Si, $^{51}$V, $^{89}$Y, $^{124}$Sn, $^{208}$Pb and $^{209}$Bi targets at various energies were also utilized for which the global OMP and SPP calculations were performed, and the reaction cross sections are listed in Table~\ref{tab:table2}. The reaction cross sections were also derived from the given set of optical model potential parameters in \cite{Harphool22,Palshetkar,Yu_2010,Bodek80,signorini2002} and are listed as $\sigma_R$ (Ref) in Table~\ref{tab:table2}. As can be seen from the table, three listed reaction cross sections are \textcolor{black} {in reasonable agreement with each other at all the energies except for the energies below the barrier where the results from SPP deviate by 30-50 \%.} 

\subsection{\label{sec:alphaenergy} $\alpha$ particle  energy and  angular distributions}
\begin{figure}
\centering
\includegraphics[width=100mm] {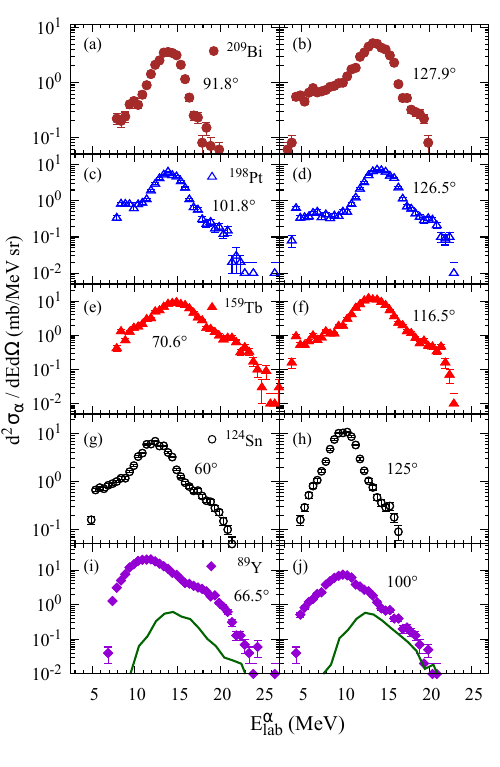}
\caption{\label{alpha_energydist} Energy spectra for $\alpha$ particles at two (grazing and backward) angles for (a,b) $^{209}$Bi, (c,d) $^{198}$Pt, (e,f) $^{159}$Tb at 35.8 MeV and (g,h) $^{124}$Sn, (i,j) $^{89}$Y at 29.7 MeV. The compound nuclear contribution from PACE2 is represented by solid line \textcolor{black}{in (i,j)}.}
\end{figure}

Energy distributions of $\alpha$ particles at two (grazing and backward) angles are shown for all the targets in Fig.\ \ref{alpha_energydist}(a-j). Statistical model calculations for compound nuclei were carried out using the code {\footnotesize PACE2} \cite{PACE} to estimate the compound nuclear contribution in the $\alpha$ spectra. The level density parameter of a = A/10 MeV$^{-1}$ was used. The default optical potentials available in {\footnotesize PACE2} for neutrons and protons \cite{PEREY76} and for $\alpha$ particles \cite{HUIZ62} emission was used. For the present data, it was found that the compound nuclear contributions for $\alpha$ particles with heavier targets (Z=83, 78, 65, and 50) were negligible (about 1\% or less), \textcolor{black}{while for medium mass $^{89}$Y target at E$_{c.m.}$ = 1.2 V$_{B}$ it was around 7\%, which is shown by solid line in Fig.\ \ref{alpha_energydist}(i-j)}.   

\begin{figure}
\centering
\includegraphics[width=90mm] {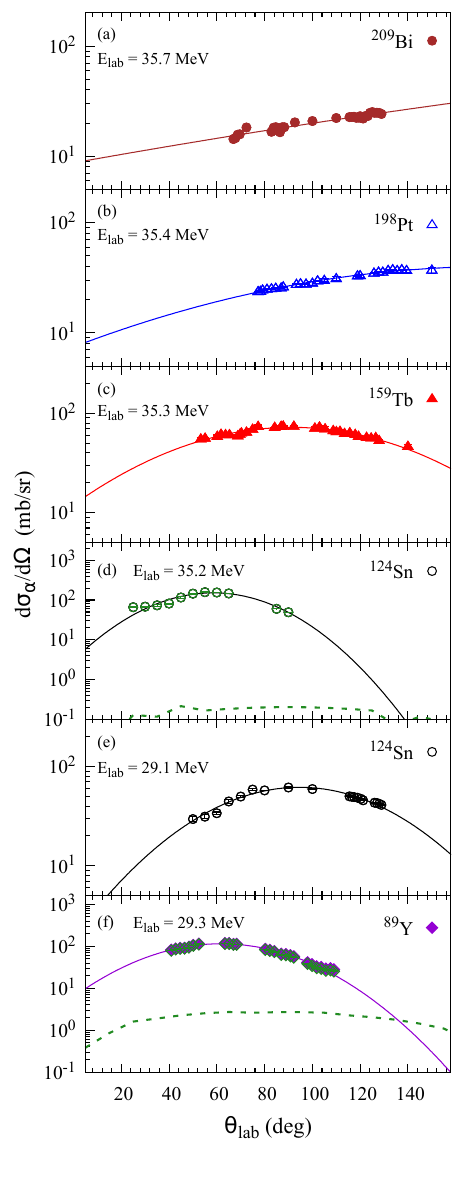}
\caption{\label{alph_ang} Angular distributions of inclusive $\alpha$ particles for reactions induced by $^9$Be projectile on different targets at beam energies (a-d) 35.8 MeV and (e-f) 29.7 MeV along with compound nuclear contribution (dashed lines) calculated using {\footnotesize PACE2}. The hollow symbols in (d,f) represents $\alpha$ angular distribution after subtracting the compound nuclear contribution. Continuous lines are guide to an eye.}
\end{figure}

Next, the energy integrated $\alpha$ particle yields were obtained at different angles, and the angular distributions at different beam energies for various targets are plotted in Fig.~\ref{alph_ang}. The errors in the cross sections are statistical only. These angular distributions are found to peak at their respective grazing angles. Also, the grazing angle is shifting towards the backward angles with the increasing atomic number of the target as expected due to the change in the Coulomb barrier.  

These angular distributions were then fitted with a Gaussian distribution, and an angle integrated total inclusive $\alpha$ cross section was obtained following the method given in Ref.\ \cite{chavi}. Negligible compound nuclear contributions were then subtracted from total inclusive $\alpha$ cross sections to get the non CF $\alpha$ cross sections and are listed as $\sigma_{\alpha}$ in Table~\ref{tab:table2}. The available non CF $\alpha$ production cross sections and the CF cross sections available with $^{51}$V, $^{89}$Y, $^{124}$Sn, $^{208}$Pb and $^{209}$Bi targets at various energies are also listed in Table~\ref{tab:table2}.  

\begin{table}
\caption{\label{tab:table2}
The measured $\alpha$ ($\sigma_{\alpha}$) and CF ($\sigma_{CF}$) cross sections are listed along with reaction cross sections obtained from Global OMP, SPP, and using potential parameters from those available in the references as (Ref.) for various targets and lab energies ($E_{lab}$) with $^9$Be projectile. Also, the reaction cross section extracted by fitting the present data is listed under ($\sigma_{R}$ (Ref.)).  The Coulomb barrier V$_{B}$ for $^9$Be projectile with $^{209}$Bi, $^{208}$Pb, $^{198}$Pt, $^{159}$Tb, $^{124}$Sn, $^{89}$Y, $^{51}$V and $^{28}$Si targets are 40.1 MeV, 39.5 MeV, 37.9 MeV, 32.9 MeV, 26.2 MeV, 21.5 MeV, 13.6 MeV, and 8.9 MeV respectively, estimated from Ref.\ \cite{NRV}.
(see text for more details) }
\begin{tabular}{cccccccccc}
\hline \hline
&&&&
&&&& \\
Targets&E$_{lab}$&E$_{c.m.}$/V$_{B}$&$\sigma_{\alpha}$
&$\sigma_{CF}$&$\sigma_{R}$&$\sigma_{R}$&$\sigma_{R}$&$\sigma_{R}$ - $\sigma_{CF}$\\
&&&&
&Ref.&Global OMP&SPP&\\
&(MeV)&&(mb)&(mb)
&(mb)&(mb)&(mb)&(mb)\\
&&
&&&& \\
\hline
&&&&
&&& \\
& 35.7 & 0.85 & 203 $\pm$ 47
& 0.62 $\pm$ 0.12$^b$ & \textcolor{black}{55$\pm$18} & 41 & 25 & \textcolor{black}{54$\pm$18} \\
& 36.2 & 0.90 & 74.7 $\pm$ 7.5$^a$
& 1.17 $\pm$ 0.13$^b$ &  & 52 & 34 & 51 $\pm$ 0.13 \\
& 37.2 & 0.93 & 114.2 $\pm$ 11.4$^a$
& 4.74 $\pm$ 0.28$^b$ & 116$^d$ & 79 & 57 & 111 $\pm$ 0.28 \\
& 38.3 & 0.96 & 229.3 $\pm$ 23$^a$
& 14.91 $\pm$ 0.42$^b$ & 138$^d$ & 132 & 98 & 123 $\pm$ 0.42 \\
& 39.2 & 0.98 & 242.5 $\pm$ 24$^a$
& 27.41 $\pm$ 1.11$^b$ & 189$^d$ & 175 & 148 & 162 $\pm$ 1.11 \\
& 40.3 & 1.01 & 282.3 $\pm$ 28$^a$
& 57.34 $\pm$ 0.57$^b$ & 285$^d$ & 254 & 227 & 228 $\pm$ 0.57 \\
& 41.3 & 1.03 & 419.9 $\pm$ 42$^a$
& 103.4 $\pm$ 0.74$^b$ & 382$^d$ & 342 & 314 & 279  $\pm$ 0.74 \\
$^{209}$Bi & 42.3 & 1.06 & 509.1 $\pm$ 51$^a$
& 202.3 $\pm$ 4.13$^b$ & 492$^d$ & 438 & 407 & 290 $\pm$ 4.13 \\
& 43.3 & 1.08 & 509.9 $\pm$ 51$^a$
& 250.7 $\pm$ 6.05$^b$ &  & 537 & 502 & 286 $\pm$ 6.05 \\
& 44.3 & 1.11 & 532.2 $\pm$ 53$^a$
& 236 $\pm$ 7$^c$ & 714$^d$ & 635 & 595 & 478 $\pm$ 7 \\
& 45.3 & 1.13 & 504.8 $\pm$ 51$^a$
& 347.2 $\pm$ 3.78$^b$ &  & 731 & 684 & 384 $\pm$ 3.78 \\
& 46.3 & 1.16 & 423.5 $\pm$ 42$^a$
& 374.4 $\pm$ 9.1$^b$ & 832$^e$ & 823 & 770 & 458 $\pm$ 9.1 \\
& 47.2 & 1.18 & 460.4 $\pm$ 46$^a$
& 452.9 $\pm$ 6.34$^b$ &  & 903 & 845 & 450 $\pm$ 6.34 \\
& 48.3 & 1.21 & 550.7 $\pm$ 55$^a$
& 521.2 $\pm$ 8.03$^b$ & 1024$^e$ & 995 & 931 & 503 $\pm$ 8.03 \\
&&&
&&&& \\
& 35.9 & 0.87 & 166 $\pm$ 8$^f$
& 0.56 $\pm$ 0.22$^g$ &  & 54 & 36 & 53 $\pm$ 0.22 \\
& 38 & 0.92 & 207 $\pm$ 13$^f$
& 9.5 $\pm$ 0.9$^g$ & 156$^d$ & 129 & 105 & 147 $\pm$ 0.9 \\
& 40 & 0.97 & 309 $\pm$ 7$^f$
& 55.6 $\pm$ 2.1$^g$ & 313$^d$ & 265 & 240 & 257 $\pm$ 2.1\\
& 42 & 1.02 & 437 $\pm$ 6$^f$
& 142.5 $\pm$ 4$^g$ & 513$^d$ & 452 & 424 & 371 $\pm$ 4 \\
$^{208}$Pb & 44.1 & 1.07 & 445 $\pm$ 10$^f$
& 243.7 $\pm$ 4.9$^g$ & 722$^d$ & 660 & 612 & 478 $\pm$ 4.9 \\
& 46.1 & 1.12 & 413 $\pm$ 10$^f$
& 351.4 $\pm$ 6.6$^g$ &  & 849 & 797 & 498 $\pm$ 6.6\\
& 48 & 1.16 & 515 $\pm$ 10$^f$
& 453.8 $\pm$ 8.6$^g$ &  & 1016 & 949 & 562 $\pm$ 8.6 \\
& 50.2 & 1.21 & 585 $\pm$ 10$^f$
& 570 $\pm$ 12$^g$ & 1263$^d$ & 1188 & 1096 & 693 $\pm$ 12 \\
&&&&&
&&&& \\
$^{198}$Pt & 35.4 & 0.89 & 264 $\pm$ 4.1
& & \textcolor{black}{66$\pm$14} & 81 & 59 &   \\
&&&&&
&&&& \\
$^{159}$Tb & 35.3 & 1.02 & 609 $\pm$ 7.8
& & \textcolor{black}{447$\pm$50} & 390 & 377 &  \\
&&&&&
&&&& \\
$^{124}$Sn & 29.1 & 1.03 & 460 $\pm$ 3.7
& 129.3 $\pm$ 5.3$^h$ & \textcolor{black}{425$\pm$31} & 396 & 390 & \textcolor{black}{296 $\pm$ 31} \\
& 35.2 & 1.25 & 663 $\pm$ 16
& 517 $\pm$ 11$^h$ & \textcolor{black}{1037$\pm$15} & 1018 & 988 & \textcolor{black}{520 $\pm$ 19} \\
&&&
&&&& \\
& 22.7 & 0.96 & 169.9 $\pm$ 60.9$^i$
& 33.2 $\pm$ 1.9$^j$ & 124 $\pm$ 18$^i$ & 173 & 153 & 91 $\pm$ 18 \\
& 24.7 & 1.04 & 377.8 $\pm$ 129.2$^i$
& 132 $\pm$ 7$^j$ & 293 $\pm$ 27$^i$ & 372 & 364 & 161 $\pm$ 28 \\
& 26.7 & 1.13 & 513.3 $\pm$ 49.5$^i$
& 265 $\pm$ 14$^j$ & 583 $\pm$ 35$^i$ & 591 & 586 & 318 $\pm$ 38 \\
$^{89}$Y& 28.7 & 1.21 & 533.6 $\pm$ 38.5$^i$
& 361 $\pm$ 20$^j$ & 716 $\pm$ 27$^i$ & 792 & 782 & 355 $\pm$ 34 \\
& 29.3 & 1.24 & 566 $\pm$ 4.8
& 495 $\pm$ 35$^i$ & \textcolor{black}{813$\pm$78} & 844 & 832 & \textcolor{black}{318 $\pm$ 85} \\
& 33.2 & 1.40 & 604.7 $\pm$ 31.19$^i$
& 559 $\pm$ 32$^j$ & 1156 $\pm$ 81$^i$ & 1172 & 1131 & 597 $\pm$ 87\\
&&&
&&&& \\
$^{51}$V & 25.3 & 1.58 & 460 $\pm$ 58$^k$ & & 1070$^k$ & 1175 & 1147 &\\
& 27.3 & 1.70 & 576 $\pm$ 45$^k$
& & 1293$^k$ & 1303 & 1263 & \\
&&&
&&&& \\
$^{28}$Si & 14 & 1.19 & 330 $\pm$ 50$^l$ & 310 $\pm$ 31$^l$ & 580 $\pm$ 29$^l$ & 535 & 509 & 270 $\pm$ 42\\
& 20 & 1.69 & 390 $\pm$ 40$^l$
& 680 $\pm$ 68$^l$ & 1140 $\pm$ 57$^l$ & 1084 & 1058 & 460 $\pm$ 89\\
&&&&
&&& \\
\hline \hline
\end{tabular}

* The measured $\sigma_{R}$ (Ref.), $\alpha$ ($\sigma_{\alpha}$) and CF ($\sigma_{CF}$) cross sections are from $^a${\cite{Sig04}}, $^b${\cite{signo99}}, $^c${\cite{Dasgupta}}, $^d${\cite{Yu_2010}}, $^e${\cite{signorini2002}}, $^f${\cite{woolli01}}, $^g${\cite{Dasgupta03}}, $^h${\cite{ParkarSn}}, $^i${\cite{Palshetkar}}, $^j${\cite{CFPalshetkar}}, $^k${\cite{Harphool22}},$^l${\cite{Bodek80}} 
\end{table}


\begin{figure}
\centering
\includegraphics[width=100mm] {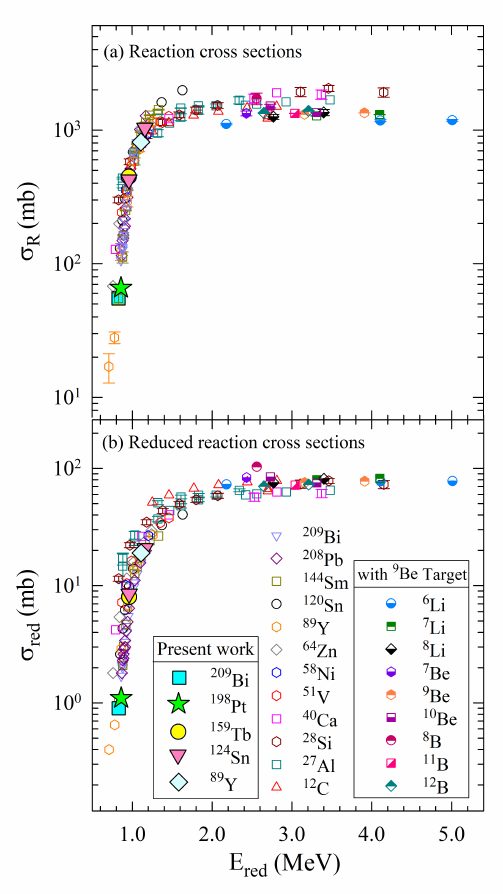}
\caption{\label{Fig_5} (a) Reaction cross sections $\sigma_{R}$ and (b) Reduced reaction cross sections $\sigma_{red}$ as a function of reduced energy ($E_{red}$) for heavy and medium-mass targets with $^9$Be projectile and for light mass projectiles with $^9$Be target. The filled data represents the present work and the unfilled/semi-filled data are from Refs.\cite{Umbelino22, Harphool22, Arazi18, Morales16, Palshetkar, Zamora11, Di10, Yu_2010, GomesSm209, Benjamim07, GomesSm06, GVM05,  PRS05, Gomes04,Signorini2000, JARCZYK81, Bodek80, JsEck80, Balzer77, DP77}}
\end{figure}


\section{Systematic study of reaction and $\alpha$ production cross sections} \label{sec:systematics}
In the following subsections, the systematics of reaction cross sections, inclusive $\alpha$ cross sections, and the ratios of inclusive $\alpha$ and measured CF cross sections to reaction cross sections are shown. In order to remove \textcolor{black}{the effect of the barrier height}, the x-axis is chosen as reduced energy E$_{red}$ = E$_{c.m.}$ ($A_{P}$$^{1/3}$ + $A_{T}$$^{1/3}$)/($Z_{P}Z_{T}$) \textcolor{black}{as used in Ref.\ \cite{Gomes05a}} where $Z_{P}$ $(Z_{T})$ and $A_{P}$ $(A_{T})$ are the atomic numbers and atomic masses of the projectile (target), respectively.

\subsection{Reaction cross sections} \label{sec:sysreaction}
The reaction cross sections obtained from the optical model calculations for the present work, along with those reported in the literature involving the $^9$Be nucleus \cite{Umbelino22, Harphool22, Arazi18, Morales16, Palshetkar, Zamora11, Di10, Yu_2010, GomesSm209, Benjamim07, GomesSm06, GVM05, PRS05, Gomes04,Signorini2000, JARCZYK81, Bodek80, JsEck80, Balzer77, DP77} are shown in Fig.\ \ref{Fig_5} (a). It is found that the reaction cross sections ($\sigma_{R}$) involving $^9$Be nucleus and nuclei lighter than $^{12}$C nucleus show a slightly different trend from the other nuclei as seen in Fig.~\ref{Fig_5} (a). To understand this, these reaction cross sections were reduced as $\sigma_{red}$ = $\sigma_{R}$/($A_{P}$$^{1/3}$ + $A_{T}$$^{1/3}$)$^2$ \cite{Gomes05a} to exclude the effects due to geometrical size as shown in Fig.~\ref{Fig_5} (b) which follows a universal trend.  

\subsection{\label{sec:alphaproduction}$\alpha$ production cross sections}

\begin{figure*} 
\centering
\includegraphics {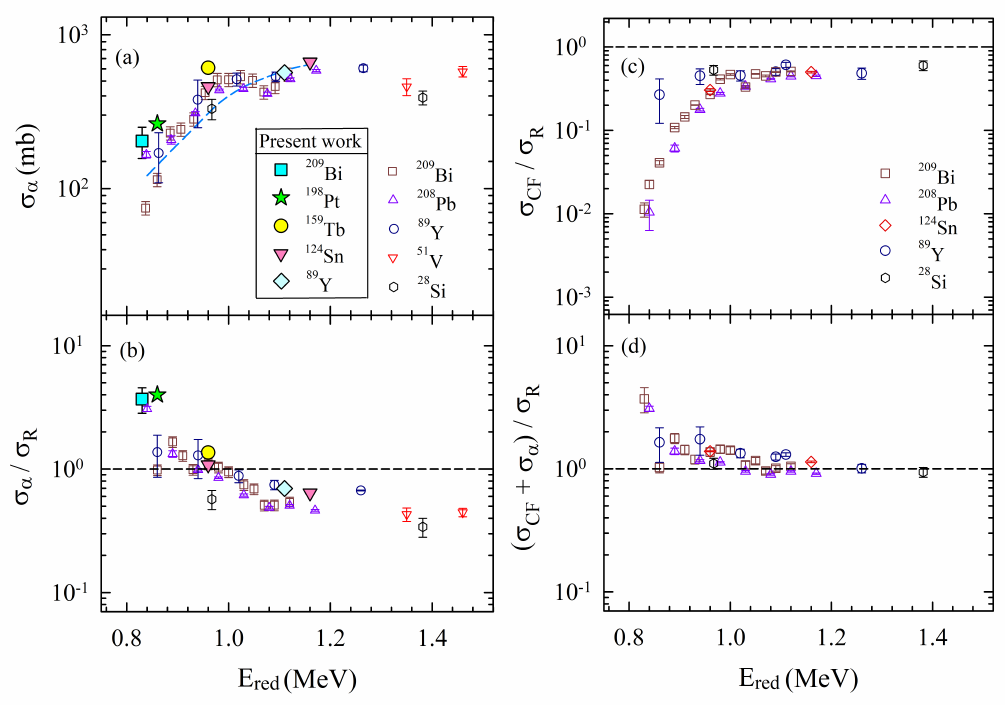}
\caption{\label{Fig_6} (a) Systematical behaviour of inclusive $\alpha$ cross sections for different targets with $^9$Be projectile as a function of reduced energy. The filled data represents present work and the unfilled data are from Ref.\cite{Harphool22,Palshetkar,Bodek80,woolli01,Sig04}. The dashed line shows $\sigma_{R}-\sigma_{CF}$ data for $^9$Be+$^{208}$Pb system. Ratio of (b) $\sigma_{\alpha}$ to $\sigma_{R}$, (c) $\sigma_{CF}$ to $\sigma_{R}$ and (d) ($\sigma_{CF}$+$\sigma_{\alpha}$) to $\sigma_{R}$, as a function of reduced energy ($E_{red}$) for various targets with $^9$Be projectile. List of data used is given in Table~\ref{tab:table2}. (see text for details)}
\end{figure*}

The inclusive $\alpha$ cross sections with the $^9$Be projectile presented in Fig.~\ref{Fig_6} (a) are from the present work and those available in the literature~\cite{Harphool22,Palshetkar,Bodek80,woolli01,Sig04}. A systematic universal behaviour in $\alpha$ cross sections is seen, similar to those observed with $^{6,7}$Li projectiles~\cite{Pakou03, Pakou05, Souz09, Kumawat10, Santra12, Pandit17, chavi, VVP_EPJA23}.  To understand this trend, the difference between reaction cross sections $\sigma_{R}$ and measured CF cross sections $\sigma_{CF}$ for the $^9$Be+$^{208}$Pb system (as a representative case) are compared and found to give a good description with the measured inclusive $\alpha$ cross sections as shown in Fig.~\ref{Fig_6} (a). Same is observed with other targets ($^{209}$Bi, $^{89}$Y and $^{124}$Sn). The $\sigma_{R}$ and $\sigma_{CF}$ are listed in Table~\ref{tab:table2}.
From the ratio of $\alpha$ \textcolor{black}{cross sections} to reaction cross sections (Fig.\ \ref{Fig_6}(b)), it is observed that the $\alpha$ cross sections are more than 100\% of the reaction cross sections at \textcolor{black}{sub-barrier} energies, which then decreases up to 40\% at above-barrier energies. At the \textcolor{black}{sub-barrier} energies, it can be interpreted that the contribution to the inclusive $\alpha$ from the reaction channels e.g. 1n-transfer, breakup of $^9$Be leading to two $\alpha$ particles are dominant over the reaction channels e.g. $\alpha$/$^5$He capture, which leads to only one $\alpha$ particle in the outgoing channel. Similar conclusions were drawn in Ref. \cite{Prasanna22} in the study of $^9$Be+$^{169}$Tm,$^{181}$Ta,$^{187}$Re and $^{197}$Au systems. As a result, at \textcolor{black}{sub-barrier} energies, the $\alpha$ production cross section is larger than the reaction cross section. 


\begin{figure} 
\centering
\includegraphics[width=120mm,trim=2.3cm 19.2cm 6.6cm 0.8cm, clip=true] {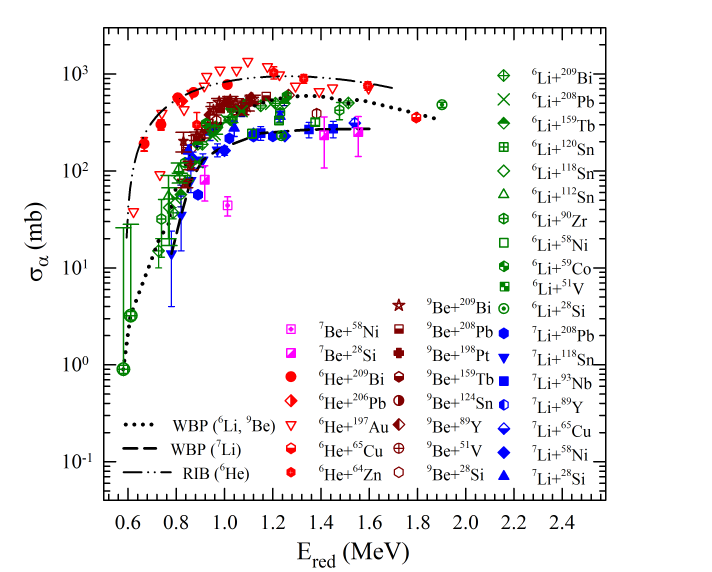}
\caption{\label{alpha_syst_WBPonly} Systematic comparison of inclusive $\alpha$ cross sections due to non-CF processes for different targets with $^{6,7}$Li \cite{Pakou03,Pakou05,Shri06,Souz09,Kumawat10,Santra12,Pand16,Chatto16,PFEIFFER1973,HUGI81,Ost1835,Signo2001,Pradhan13,Pandit19}, $^9$Be \cite{Harphool22,Palshetkar,Bodek80,woolli01,Sig04}, $^6$He \cite{Navin2004,Chatterjee2008,DiP04,Scuderi11,YuE07,Standylo13,Aguilera2000,Kolata2002} and $^7$Be \cite{Mazzocco15,Sgouros16} projectiles as a function of reduced energy. (see text for details)}
\end{figure}

\begin{figure} 
\centering
\includegraphics[width=102mm,trim=0.9cm 5.2cm 14.8cm 5.6cm, clip=true] {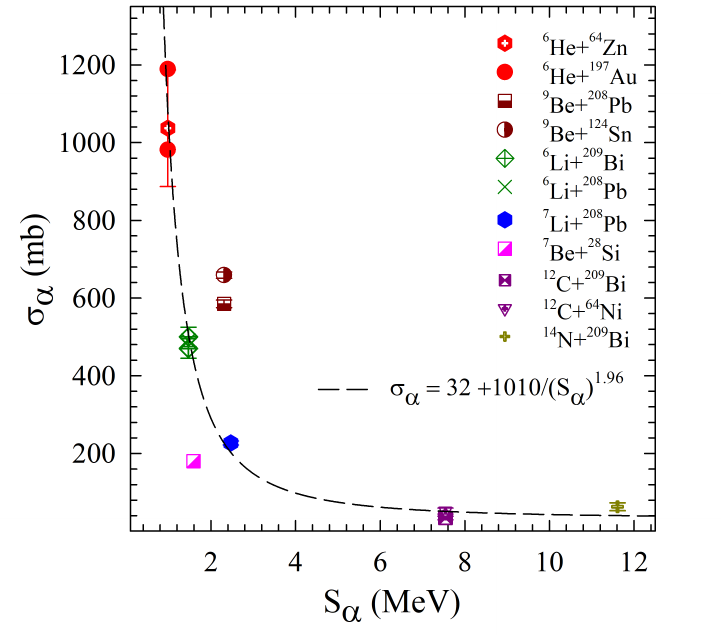}
\caption{\label{Fig9} Non-CF inclusive $\alpha$ cross section for with various targets for projectiles $^6$He~\cite{DiP04,YuE07},  $^{6,7}$Li~\cite{Santra12,Signo2001}, $^7$Be~\cite{Sgouros16}, $^9$Be~\cite{woolli01}, $^{12}$C \cite{JIN12C,Zhang88}, and $^{14}$N \cite{Nomura_14N} around 1.2 Vb as a function of $\alpha$ separation energies for projectiles  (S$_\alpha$) is shown. The line is fit to the data (see text for details).}
\end{figure}

To compare the individual and cumulative contributions of CF and $\alpha$ cross sections in reaction cross sections, we have plotted the fraction of $\alpha$, CF, and cumulative CF+$\alpha$ cross sections over reaction cross sections in Fig.\ \ref{Fig_6}(b), Fig.\ \ref{Fig_6}(c) and Fig.\ \ref{Fig_6}(d) respectively. The data available in the literature for CF and $\alpha$ cross sections with various targets are given in Table~\ref{tab:table2}. 
As shown in Fig.\ \ref{Fig_6}(c), the ratio of CF to reaction cross sections increases with energy and saturates around 50-60\% at energies above the barrier for all targets. 
According to Fig.\ \ref{Fig_6}(b) and Fig.\ \ref{Fig_6}(c), it can be interpreted that the inclusive $\alpha$ is dominant at \textcolor{black}{sub-barrier} energies and CF is dominant at above-barrier energies. The sum of these two cross sections (CF+$\alpha$), shown in Fig.\ \ref{Fig_6}(d), is found to be greater than the reaction cross sections, owing primarily to the double $\alpha$ events from the 1n-transfer and breakup of the $^9$Be projectile, as discussed in Sec.\ \ref{sec:alphaenergy} and Sec.\ \ref{sec:alphaproduction}. 

Inclusive $\alpha$ cross sections with other weakly bound stable and unstable projectiles are also compared in Fig.\
\ref{alpha_syst_WBPonly}. Lines are shown \textcolor{black}{to guide the} eye. Non CF inclusive $\alpha$ production data available with $^{6,7}$Li \cite{Pakou03,Pakou05,Shri06,Souz09,Kumawat10,Santra12,Pand16,Chatto16,PFEIFFER1973,HUGI81,Ost1835,Signo2001,Pradhan13,Pandit19}, $^9$Be \cite{Harphool22,Palshetkar,Bodek80,woolli01,Sig04}, $^6$He \cite{Navin2004,Chatterjee2008,DiP04,Scuderi11,YuE07,Standylo13,Aguilera2000,Kolata2002} and $^7$Be \cite{Mazzocco15,Sgouros16} projectiles are utilised for making this systematic study. As can be seen, an increase in $\sigma_{\alpha}$ with incident energy is observed for all the projectiles. There is a distinction between $^6$He compared with $^{6,7}$Li and $^9$Be induced reactions. \textcolor{black}{Among $^{6,7}$Li and $^9$Be projectiles, the $\alpha$ production with} $^6$Li and $^9$Be \textcolor{black}{is} higher than $^7$Li \textcolor{black}{projectile}. Surprisingly limited data available with $^7$Be projectile on $^{28}$Si and $^{58}$Ni targets are similar to $^7$Li projectile. Further investigation is required to understand the systematic trend. 

In order to investigate further, the dependence of non-CF $\alpha$ cross sections at around 1.2 V$_b$ for $^6$He~\cite{DiP04,YuE07}, $^{6,7}$Li~\cite{Santra12,Signo2001}, $^7$Be~\cite{Sgouros16}, $^9$Be~\cite{woolli01}, and $^{12}$C \cite{JIN12C,Zhang88} $^{14}$N \cite{Nomura_14N}  projectiles with various targets on $\alpha$ separation energies $S_{\alpha}$ of the projectiles are shown in Fig.\ \ref{Fig9}. Earlier, $\alpha$ production cross sections was shown to correlate with both $\alpha$ and neutron separation energies~\cite{VVP_EPJA23,kolata01}. As can be seen, all the systems, except $^{7,9}$Be, show a systematic behaviour. The line represents a fit to the data (excluding $^{7,9}$Be data) with the function a+(b/x$^c$), \textcolor{black}{where a=32, b=1010, and c= 1.96}. While the $^9$Be shows enhancement in $\alpha$ yields, $^7$Be exhibits suppression with respect to the systematic behaviour. The breakup and 1n-transfer of $^9$Be projectiles lead to double $\alpha$ events. Therefore, it might be a reason for enhanced $\alpha$ yields. In fact, it is observed that the measured cross sections of $\alpha$ producing channels (Direct+Incomplete fusion) for $^9$Be projectile with $^{238}$U~\cite{Raabe06} and $^{197}$Au ~\cite{malika2021} targets at around 1.2 V$_b$ are about 263-340 mb, which are close to the value according to the systematics of Fig.\ \ref{Fig9}. There are very few measurements using the $^7$Be projectile. In the present study, the $^7$Be data point was obtained by interpolating the available experimental data~\cite{Sgouros16}, and was found to be significantly lower as compared to the systematics.  However, the sum of measured cross sections of Direct+Incomplete fusion ($\alpha$ producing channels ) at around 1.2 V$_b$ is about 344 mb for $^7$Be+$^{238}$U~\cite{Raabe06}, which is in good agreement with the systematics of Fig.\ \ref{Fig9}. Therefore, more studies are required to understand the behaviour of Be projectiles.



\section{Summary} \label{sec:summary}

The elastic scattering and inclusive $\alpha$ particle angular distributions have been measured with $^9$Be projectile on $^{89}$Y, $^{124}$Sn, $^{159}$Tb, $^{198}$Pt, and $^{209}$Bi targets at energies around the Coulomb barrier. To obtain the reaction cross section, the elastic scattering angular distribution has been compared with the optical model calculations using both, phenomenological global optical model potential (Global OMP) and microscopic S$\tilde{a}$o Paulo potentials (SPP). The systematic study of reaction and inclusive $\alpha$ cross sections for the present work and those reported in the literature were found to follow a universal behavior. It is observed that for energies below the barrier, the $\alpha$ production cross sections are larger than the reaction cross section mainly due to the 1n-transfer and breakup processes which contribute to two $\alpha$ particles. At \textcolor{black}{sub-barrier} energies inclusive $\alpha$ is dominant, and for above-barrier, CF is dominant, and the sum of these two processes (CF+$\alpha$) almost explains the reaction cross sections at all the energies over a wide range of target mass.  The $\alpha$ production cross sections with weakly bound projectiles $^6$He, $^{6,7}$Li and $^{7,9}$Be have also been compared. It is observed that the $\alpha$ production with $^6$He projectile is larger than others at all the energies. Among $^{6,7}$Li and $^9$Be projectiles, $\alpha$ production with $^{6}$Li and $^9$Be projectile are similar and larger than $^{7}$Li. Further, the non-CF $\alpha$ cross sections were found to correlate well with separation energies $S_{\alpha}$ of the projectiles.

\section{Acknowledgments}
We would like to thank the Pelletron Linac Facility staff of BARC-TIFR, Mumbai, India, for smooth operation during the experi-
ment. We are also thankful to Mr. P. Patale for his help during the experiment. We also thank target lab staff of TIFR for their help in
the target preparation. One of the authors (S.K.) acknowledges the fellowship through the CSIR-UGC India through NET-JRF scheme.
S.D. acknowledges the funding from DST-INSPIRE Fellowship scheme of Govt. of India.


\end{document}